 \documentclass[preprint,review,12pt,sort&compress]{elsarticle}
\usepackage{lipsum}

\let\OLDthebibliography\thebibliography
\renewcommand\thebibliography[1]{
  \OLDthebibliography{#1}
  \setlength{\parskip}{0pt}
  \setlength{\itemsep}{0pt plus 0.1ex}
}
\usepackage{enumitem}
\usepackage[margin=1.0 in]{geometry}
\usepackage{multirow}
\newcommand{\B}{Ba\v zant}  

 \newcommand{\bc}{\begin{center}}
 \newcommand{\ec}{\end{center}}
                   \newcommand{\bfr}{\begin{flushright}}
                   \newcommand{\efr}{\end{flushright}}
   \newcommand{\ii}{\item}
     \newcommand{\be}{\begin{enumerate}}
     \newcommand{\ee}{\end{enumerate}}
        \newcommand{\bi}{\begin{itemize}}
        \newcommand{\ei}{\end{itemize}}
            \newcommand{\bd}{\begin{description}}
            \newcommand{\ed}{\end{description}}
                \newcommand{\beq}{\begin{equation}}
                \newcommand{\eeq}{\end{equation}}
                  \newcommand{\bea}{\begin{eqnarray}}
                  \newcommand{\eea}{\end{eqnarray}}

      \newcommand{\bfi}{\begin{figure}}
      \newcommand{\efi}{\end{figure}}
\newcommand{\bay}{\begin{array}{l}}
\newcommand{\eay}{\end{array}}
            
    \newcommand{\eps}{\epsilon}

\usepackage{amsmath}
\usepackage{breqn}
\usepackage{textcomp}

\usepackage{amssymb}

\journal{Composites Science and Technology}

\begin{document}

\begin{titlepage}
\clearpage\thispagestyle{empty}
\noindent
\hrulefill
\begin{figure}[h!]
\centering
\includegraphics[width=2 in]{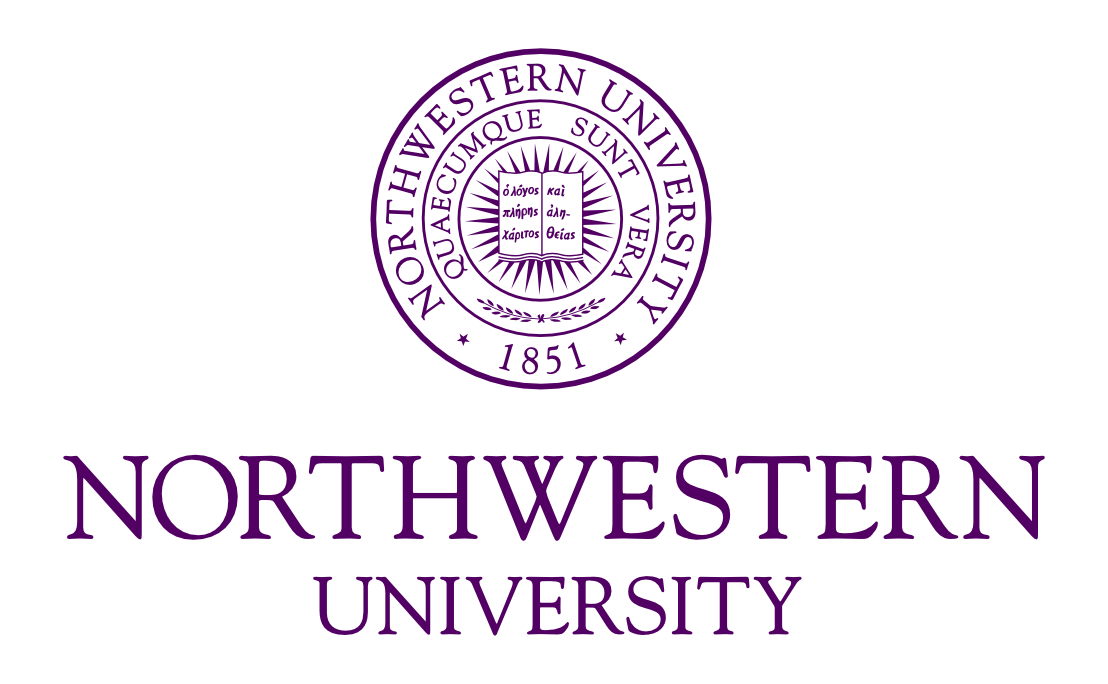}
\end{figure}
\begin{center}
{
{\bf Center for Sustainable Engineering of Geological and
Infrastructure Materials (SEGIM)} \\ [0.1in]
Department of Civil and Environmental Engineering \\ [0.1in]
McCormick School of Engineering and Applied Science \\ [0.1in]
Evanston, Illinois 60208, USA
}
\end{center} 
\hrulefill \\ \vskip 2mm
\vskip 0.5in
\begin{center}
{\large {\bf \uppercase{Experimental and Numerical Investigation of Intra-Laminar Energy Dissipation and Size Effect in Two-Dimensional Textile Composites}}}\\[0.5in]
{\large {\sc Marco Salviato, Kedar Kirane, Shiva Esna Ashari, Zden\v ek Ba\v zant, Gianluca Cusatis}}\\[0.75in]
{\sf \bf SEGIM INTERNAL REPORT No. 16-05/707E}\\[0.75in]
\end{center}
\noindent {\footnotesize {{\em Submitted to Composites Science and Technology \hfill May 2016} }}
\end{titlepage}

\newpage
\clearpage \pagestyle{plain} \setcounter{page}{1}

\begin{frontmatter}



\title{Experimental and Numerical Investigation of Intra-Laminar Energy Dissipation and Size Effect in Two-Dimensional Textile Composites}


\author[label1]{Marco Salviato}
\author[label4]{Kedar Kirane}
\author[label4]{Shiva Esna Ashari}
\author[label4]{Zden\v ek Ba\v zant}
\author[label4]{Gianluca Cusatis\corref{cor1}}
\address[label1]{ Department of Aeronautics and Astronautics, University of Washington, Seattle, WA 98195, USA}
\address[label4]{Department of Civil and Environmental Engineering, Northwestern University, Evanston, IL 60208, USA}

\cortext[cor1]{Corresponding Author, \ead{g-cusatis@northwestern.edu}}

\address{}

\begin{abstract}
\linespread{1}\selectfont

Design of large composite structures requires understanding the scaling of their mechanical properties, an aspect often overlooked in the literature on composites.

This contribution analyzes, experimentally and numerically, the intra-laminar size effect of textile composite structures. Test results of geometrically similar Single Edge Notched specimens made of $[0^{\circ}]_8$ epoxy/carbon twill 2$\times$2 laminates are reported. Results show that the nominal strength decreases with increasing specimen size and that the experimental data can be fitted well by Ba\v zant's size effect law, allowing an accurate identification of the intra-laminar fracture energy of the material, $G_f$.

The importance of an accurate estimation of $G_f$ in situations where intra-laminar fracturing is the main energy dissipation mechanism is clarified by studying numerically its effect on crashworthiness of composite tubes.
Simulations demonstrate that, for the analyzed geometry, a decrease of the fracture energy to 50$\%$ of the measured value corresponds to an almost 42$\%$ decrease in plateau crushing load. Further, assuming a vertical stress drop after the peak, a typical assumption of strength-based constitutive laws implemented in most commercial Finite Element codes, results in an strength underestimation of the order of 70$\%$.

The main conclusion of this study is that measuring accurately fracture energy and modeling correctly the fracturing behavior of textile composites, including their \emph{quasi-brittleness}, is key. This can be accomplished neither by strength- or strain-based approaches, which neglect size effect, nor by LEFM which does not account for the finiteness of the Fracture Process Zone.

\end{abstract}


\end{frontmatter}

\linespread{1.4}\selectfont
\section{Introduction}
\label{}
Thanks to their outstanding specific mechanical properties, the engineering use of textile composites is becoming broader and broader. Current applications include land, marine and air transportation, wind and tidal energy production, and blast protection of civil infrastructures and vehicles \cite{Cho92,DanIsh92,BogPas96}. However, design of large composite structures requires capturing the scaling of their mechanical properties, an aspect often overlooked in the literature on composites. This can be achieved only by abandoning the current design paradigm, which relies on strength-based approaches incapable of predicting any scaling, and acknowledging the \emph{quasibrittle} character of these materials.

Due to the complex mesostructure characterizing textile composites (and other quasibrittle materials such as concrete, nanocomposites, ceramics, rocks, sea ice, and many bio-materials, just to mention a few), the size of the non-linear Fracture Process Zone (FPZ) occurring in the presence of a large stress-free crack is usually not negligible \cite{BazPla98,GreWisHal07,MolIarKim06}. The stress field along the FPZ is nonuniform and decreases with crack opening, due to discontinuous cracking, crack bridging by fibers, and frictional pullout of inhomogeneities. As a consequence, the fracturing behavior and, most importantly, the energetic size effect associated with the given structural geometry, cannot be described by means of the classical Linear Elastic Fracture Mechanics (LEFM). To capture the effects of a finite, non-negligible FPZ, the introduction of a characteristic (finite) length scale related to the fracture energy and the strength of the material is necessary \cite{BazDanLi96,BazPla98}. However, estimating accurately these material properties is far from easy because the fracture tests usually exhibit an extreme snap-back at peak load, with a loss of stability \cite{BazDanLi96, CusBegBaz08}.

A possible way to overcome these issues is size effect testing \cite{BazDanLi96,BazPla98}. This study proposes an experimental and numerical investigation on the efficacy of the intra-laminar size effect testing to characterize the fracturing behavior of textile composite.
It is worth remarking here that the size effect method of measuring the fracture properties is easier to implement than other methods because only peak load measurements are necessary. The post-peak behavior, crack tip displacement measurement, and optical measurement of the crack tip location are not needed, and even a soft testing machine without servo-control can be used.

\section{Test description}

\subsection{Materials}

 Experiments were conducted on specimens manufactured by compression molding. A Bisphenol A diglycidyl ether (DGEBA)-based epoxy resin was chosen as polymer matrix whereas the reinforcement was provided by a twill 2$\times$2 fabric made of carbon fibers. The material was characterized following the ASTM standard procedures \cite{ASTMD3039} testing [$0^{\circ}$]$_8$ and [$45^{\circ}$]$_8$ coupons under uniaxial tension. The results of this characterization are listed in Table \ref{T1}.

\subsection{Specimen characteristics}

Following \B~\emph{et al.} \cite{BazDanLi96,BazPla98}, intra-laminar size effect tests were conducted on single-edge-notched tension (SENT) specimens (see Figure \ref{f1}), using a [$0^{\circ}$]$_8$ lay-up with a constant thickness of approximately 1.9 mm. The SENT specimens were preferred to Double-Edge Notched Tension (DENT) specimens, for which two cracks typically initiate at the notch tips but ultimately only one of the two cracks can propagate, causing the response to be asymmetric \cite{BazTab92}.

Specimens of three sizes (three for each size), geometrically scaled in two-dimension (see Table \ref{T2}) as 1:2:4, were tested. The first half of the notch was made by means of a diamond coated bend saw which provided a width of roughly 1 mm whereas the second half was made using a diamond-coated miniature blade thanks to which a width of 0.2 mm was obtained in all cases (Figure \ref{f2}). Accordingly, the resulting crack tip radius was 0.1 mm, about 70 times smaller than the size of a Representative Unit Cell (RUC) of the material. It is worth noting that the sawing action of the blade prevented the formation of a Fracture Process Zone (FPZ) before running the tests contrarily to common pre-fracturing procedures \cite{ASTM5045}.

All the specimens were prepared with 38 mm long glass/epoxy tabs for gripping purposes. The tab length (grip constraint) was not scaled because it has no appreciable effect on the stored energy and because fracture always occurs away from the grips.

The top surface of all the SENT specimens was treated to allow Digital Image Correlation (DIC) analysis. A thin layer of white paint was deposited on a $D \times D$ area embedding the crack. Then, black speckles of average size 0.01 mm were spray-painted on the surface after drying.

\subsection{Testing}

The tests were performed on a closed-loop servohydraulic MTS machine with $89$ kN capacity and at constant crosshead rate (stroke control). The rate was adjusted for the different sizes to achieve roughly the same strain rate of 0.2 percent/min in the gage section. With such settings, the test lasted no longer than approximately 10 min for all specimens.

Stroke, force, and loading time were recorded with a sampling frequency of 10 Hz. A DIC system from Correlated Solutions \cite{Correlated} composed by a 5 MP digital camera and a workstation for image postprocessing was used to measure the displacement field in the specimen with an acquisition frequency of 1 Hz.

\section{Experimental results}

After the completion of the experiments, the load and displacement data were analyzed. Figure \ref{f3}a shows, for the various sizes,
the typical load-displacement plots reported. It is worth noting that, for the largest specimen size, these curves are almost linear up to failure, which is an indication of pronounced brittle behavior. Conversely, a significant nonlinear segment before the peak stress indicates hardening inelastic behavior and reduced brittleness (or higher ductility) for the smallest specimen sizes.


After reaching the peak load, the specimens exhibited snap-back instability for all investigated sizes. As a consequence, the failures were catastrophic (dynamic), and occurred shortly after the peak load. Damage consisting of microcracks in layers, delamination between layers before peak load and tow breakage and pull-out was observed in the tests. Figure \ref{f3}b shows the
typical appearance of the specimens after failure and the test results for the notched specimens are summarized in Table \ref{T3}. The table also reports the specimen nominal strength. This is defined as the average stress at failure based on the unnotched cross section, $\sigma_N=P_{\max}/Dt$.

It is worth noting that, according to strength-based criteria (such as e.g. Tsai and Wu \cite{TsaWu72} among others), the nominal strength does not depend on the structural size. However, Table \ref{T3} does show a significant decrease of $\sigma_N$ with increasing characteristic size of the specimen. It is clear that strength based criteria cannot capture this trend. However, neither can classical Linear Elastic Fracture Mechanics (LEFM) which, instead, would predict a decrease proportional to $D^{-1/2}$.

\section{Discussion}\label{Discussion}

\subsection{Analysis of intra-laminar size effect tests by Size Effect Law}

The intra-laminar size effect tests can be analyzed by means of an equivalent linear elastic fracture mechanics approach, which results in an equation, known as type II Size Effect Law (SEL) \cite{Baz84s,BazPla98}, which relates the nominal strength, $\sigma_N$, to the characteristic size of the structure, $D$.

\subsubsection{Energy release rate and size effect law for orthotropic materials}

Following Bao \emph{et al}. \cite{Bao91}, the Mode I stress intensity factor for an orthotropic material can be written as:
 \beq \label{e2}
  K_I=\sigma_N\sqrt{\pi D\alpha}~\xi\left(\alpha,\lambda^{1/4}L/D,\rho\right)
 \eeq
where $\alpha=a/D=$ dimensionless crack length, $\lambda^{1/4}L/D=$ rescaled length/width ratio, $\xi~(\alpha,\lambda^{1/4}L/D,\rho)$ is a dimensionless function accounting for geometric and elastic effects and $\rho$ and $\lambda$ are dimensionless elastic parameters defined as follows:
 \beq\label{e1}
  \rho=\frac{\sqrt{E_1E _2}} {2G_{12}} -\sqrt{\nu_{12}\nu_{21}}~, \qquad \lambda=\frac{E_2}{E_1}
 \eeq
In the previous expressions, $1$ and $2$ represent the weft and warp yarn directions, respectively (Figure \ref{f2}) and $E_1\approx E_2$, $G_{12}$, $\nu_{12}=\nu_{21}$ are the in-plane elastic constants of the textile composite.

The energy release rate can be written starting from the stress intensity factor as:
 \beq\label{e3}
  G(\alpha)=\sqrt{\frac{1+\rho}{2E_1 E_2\sqrt{\lambda}}}K_I^2
 \eeq
Now, recalling Eq. (\ref{e2}), one can write the energy release rate as a function of relative crack length as:
 \beq\label{e4}
  G\left(\alpha\right)=\frac{K_I^2}{E^*}=\frac{\sigma_N^2 D}{E^*}g\left(\alpha\right)
 \eeq
where
 \beq\label{e5}
  E^*=\sqrt{\frac{2E_1E_2\sqrt{\lambda}}{1+\rho}},\qquad g\left(\alpha\right)=\pi \alpha \left[\xi\left(\alpha,\lambda^{1/4}L/D,\rho\right)\right]^2
 \eeq
It should be noted that Eq. (\ref{e5}) defines an effective elastic modulus, $E^*$, dependent on the orthotropic properties of the composite while $g(\alpha)$ is the dimensionless energy release rate. Thanks to the foregoing expressions, the relation between the nominal strength and the structure characteristic size has now the same form as the isotropic case (see e.g. \cite{BazPla98}).
In this case, the failure condition can be written \cite{Baz84s,BazPla98} with reference to an effective crack length as:
 \beq\label{e9}
  G\left(\alpha_0+c_f/D\right)=\frac{\sigma_N^2 D}{E^*}g\left(\alpha\right)=G_f
 \eeq
where $G_f=$ initial fracture energy of the material and $c_f=$ effective FPZ length, both assumed to be material properties. It should be remarked that this equation characterizes the peak load conditions if $g^'(\alpha)>0$, i.e. only if the structure has positive geometry \cite{Baz84s,BazPla98}.

By approximating $g(\alpha)$ with its Taylor series expansion at $\alpha_0$
and retaining only up to the linear term of the expansion, one obtains:
 \beq\label{e10}
  \sigma_N=\sqrt{\frac{E^*\, G_f}{D g(\alpha_0)+c_f g'(\alpha_0)}}
 \eeq
This equation relates the nominal strength of radially scaled structures to a characteristic size, $D$ and it can be rewritten in the following form:
 \beq\label{e11}
  \sigma_N=\frac{\sigma_0}{\sqrt{1+D/D_0}}
 \eeq
where $\sigma_0=(E^*\,G_f/c_fg^'(\alpha_0))^{1/2}$; and $D_0=c_f g^'(\alpha_0)/g(\alpha_0)=$ constant, depending
on both FPZ size and specimen geometry. Contrarily to classical LEFM, Eq. (\ref{e11}) is endowed with a characteristic length scale $D_0$. This is the key to describe the transition from ductile to brittle behavior with increasing structure size.

\subsubsection{Fitting of experimental data by SEL}

The parameters of SEL, Eq. (\ref{e11}), can be determined by regression analysis of experimental data. To this aim, it is convenient to define the following:
 \beq\label{e12}
  X=D, \qquad Y=\sigma_N^{-2}
 \eeq
 \beq\label{e14}
  \sigma_0=C^{-1/2}, \qquad D_0=\frac {C} {A} = \frac {1} {A\left(Bf_t\right)^2}
 \eeq
Eq. (\ref{e11}) can now be expressed in the following form:
 \beq\label{e16}
  Y = C + AX
 \eeq

A linear regression analysis was conducted as represented in Figure \ref{f5}a and provided the following parameter estimates $A=0.305$ GPa$^{-2}$mm$^{-1}$ and $C=2.419$ GPa$^{-2}$, and from Eqs. (\ref{e14}a,b), $D_0=7.93$ mm and $  \sigma_0=643$ MPa.

The fitting of the experimental data by SEL is shown in Figure \ref{f5}b where the normalized strength, $\sigma_N/  \sigma_0$  is plotted as a function of the normalized characteristic size $D/D_0$ in double logarithmic scale. The figure shows a transition from the strength criterion (plastic limit analysis)
characterized by a horizontal asymptote, to an asymptote of slope $-1/2$, representing LEFM. The intersection of the two asymptotes corresponds
to $D = D_0$, called the \emph{transitional size}.

The experimental results in Figure \ref{f5}b clearly show that: (1) the failure of textile composite laminates containing traction-free cracks (or notches) exhibits a significant size effect; and (2) the size effect represents a gradual transition with increasing size from the strength criterion (e.g., maximum stress) to LEFM.

These conclusions ought to be taken into
account in all design situations and safety evaluations where a
large traction-free crack can grow in a stable manner prior to
failure. In particular, these conclusions are important for extrapolation
from small-scale laboratory tests to real size structures. The strength theory, which does not account
for size effect, is inadequate for these applications.

\subsection{Estimation of fracture properties from SEL}

The parameters of the size effect law, $A$ and $C$, can be directly related to $G_f$ and the effective FPZ length, $c_f$, as follows:
 \beq\label{e17}
  A=g(\alpha_0)/E^*G_f, \qquad C=c_f g'(\alpha_0)/E^*G_f
 \eeq
provided that the dimensionless functions $g(\alpha)$, $g^'(\alpha) = \mbox{d}g/\mbox{d}\alpha$, and the elastic constant $E^*$ are known.

\subsubsection{Calculation of $g(\alpha)$, $g^'(\alpha)$, and initial fracture energy}
\label{energy release}

Assuming uniform applied stress as boundary condition, Bao \emph{et al}. \cite{Bao91} showed that the effect of the gauge length becomes negligible for $\lambda^{1/4}L/D\ge2$ and the dimensionless function $\xi$ can be rewritten as:
 \beq\label{e8}
  \xi\left(\alpha,\lambda^{1/4}L/D,\rho\right)
  = F\left(\alpha\right)Y\left(\rho\right)
 \eeq
where $Y(\rho) = \left[1+0.1(\rho-1)-0.016(\rho-1)^2 + 0.002(\rho-1)^3 \right] \left(\frac{1+\rho} 2 \right)^{-1/4}$ accounts for the effects of orthotropy and $F(\alpha)$ is the same geometrical function of the relative crack length as for isotropic materials. For the assumed boundary conditions, this solution provides an error always lower than $2\%$ and it was used in \cite{BazDanLi96} to calculate $g(\alpha)$ and $g^'(\alpha)$ to study size effect in unidirectional composites. Numerical investigation performed in this study shows that such formula is also accurate within 3\% error for the geometry used in this study characterized by $\lambda^{1/4}L/D<2$. For $\alpha_0$=0.2 one obtains: $g(\alpha_0)=1.1460$ and $g^'(\alpha_0) = 11.25$. However, since the specimens are clamped at the tabs, a uniform remote displacement rather than stress might be a more realistic boundary condition.

In the absence of an analytical solution for displacement boundary conditions, the function $g\left(\alpha,\lambda^{1/4}L/D,\rho\right)$ was calculated by Finite Element Analysis in Abaqus Implicit 6.13 \cite{Abaqus13}. 8-node biquadratic plane stress quadrilateral elements (CPS8) were adopted while the quarter element technique \cite{Barsoum74} was used at the crack tip to provide accurate results. The smallest element size at the tip was about $a_0\times 10^{-5}$ leading to roughly $22,000$ elements for the whole model. A linear elastic orthotropic constitutive model was used for the simulation with material properties given in Table \ref{T1}. The $J$-integral approach \cite{Rice68} was adopted to estimate the energy release rate in the presence of a uniform displacement, $u_0$, applied in correspondence to the tabs, treated as rigid parts. Then, the related load, $P$, was computed and used to calculate the corresponding dimensionless energy release $g$ by means of Eq. (\ref{e9}).

In order to compute $g^'(\alpha)$, various dimensionless crack lengths in the close neighborhood of the initial crack value $\alpha_0=0.2$ were considered, namely $\alpha=0.1950$, $0.1975$, $0.2025$ and $0.2050$. As can be noted from Figure \ref{f7}a, linear interpolation provided a very accurate fit of the numerical data with $Y=g(\alpha)=2.8352X+ 0.0886$ and $X=\alpha$. According to this analysis one has $g(\alpha_0)=0.6556$ and $g^'(\alpha_0)=2.8352$. Since the specimens are geometrically scaled, these results apply to all the considered sizes.

From Eqs. (\ref{e17}a,b), the initial fracture energy and the effective FPZ length can be calculated: one has $  G_f = 73.7~ \mbox{N/mm}, c_f = 1.81~ \mbox{mm}$, for uniform displacement BCs; and $G_f = 130$ N/mm, $c_f = 4.07~ \mbox{mm}$ for uniformly applied remote stress. Clearly, the boundary conditions have a non-negligible effect which requires further investigation carried out hereinafter through a combined approach integrating numerical analysis and DIC measurements.

\subsection{Calibration of computational models through size effect tests} \label{size effect}

Intra-laminar size effect is a key factor for damage tolerance design of large textile composite structures, the assessment of which requires accurate fracture models. These models ought to be able to capture the size effect, so that the tests described in the foregoing sections could be used to provide data for calibration and validation. To clarify this point, the size effect tests were simulated by means of two recently proposed models for textile composites, namely the Spectral Stiffness Microplane Model (SSMM) \cite{SalAshCus15} and the Microplane Triad Model (MTM) \cite{KirSalBaz15a,KirSalBaz15b} for which the calibrated parameters are listed in Appendix.

\subsubsection{Microplane modeling of textile composites}

In the formulations used in this investigation, the constitutive laws are expressed in terms of stress and strain vectors acting on planes of several orientations within the material meso-structure, following the framework of microplane theory \cite{Baz84,BazOh85,CanBaz13}. Then, a variational principle is applied to relate the microplane stresses at the mesoscale to the continuum tensors at the macroscale. Thanks to these features, the models can easily capture various physical inelastic phenomena typical of fiber and textile composites such as matrix microcracking, micro-delamination, crack bridging, pullout, and debonding (for more details, refer to \cite{KirSalBaz15a, KirSalBaz15b, SalAshCus15}).

To ensure objective numerical results in the presence of strain localization, both formulations employ the crack band model proposed by \B~\emph{et al.} \cite{BazOh83}. In this approach, the width of the damage localization band, $w_c$, is considered as a material property. This width is also equal to the mesh size, $h_e$, which is here chosen as 2 mm. A change in the element size requires the scaling of the post-peak response of the material such that the fracture energy remain unchanged. Thanks to the crack band model, a characteristic size of the material is inherently embedded in the formulations, which is a key feature to correctly describe the transition from ductile to brittle behavior.

It is worth mentioning here that the post-peak responses of the present microplane formulations are somewhat different. In SSMM, the softening in weft or warp direction occurs in the form of an exponential decay whereas, in MTM, the first initial drop in stress is considered to be followed by a small plateau, and then by an exponential decay. In the latter case, the initial drop captures the assumed initial straightening of the fibers after matrix microcracking. In the absence of direct experimental observations on the softening response of composite materials, both types of post-peak behavior were considered in this contribution. However, both softening laws can be characterized by the initial fracture energy, $G_f$, and the total fracture energy, $G_F$ as depicted in Figure \ref{f8}a,b. The total fracture energy, $G_F$, corresponds to the total area under the stress-strain curve multiplied by the characteristic size (or width) of the crack band whereas the initial fracture energy, $G_f$, is related only to the initial part of the curve. It is worth observing that for common specimen sizes, the size effect tests provide information only on the initial fracture energy. This is because, for lab-scale structures of quasibrittle materials, the stress at peak load in the FPZ still remains in the initial steep portion of the post-peak stress-displacement curve, while the tail portion is reached only after the load is reduced substantially \cite{CusSch09, HooBaz14}.

\subsubsection{Simulation of size effect tests}

Finite element models for all three SENT coupons were built in Abaqus Explicit \cite{Abaqus13} and uniaxial tensile simulations were performed to fit the peak loads.  For all the considered sizes, the displacement field measured during the tests by DIC was applied as boundary conditions in finite element simulations. For this purpose, the relative displacement with respect to the left side of the specimens; i.e. the side that contains the crack mouth, was computed along two sections at the top and bottom of the crack plane at a distance of $D/2$ for each specimen of each size. This distance was chosen far enough from the crack plane to avoid the need for any assumption on the applied boundary conditions. Afterwards, the displacement difference between top and bottom points was calculated through the normalized experimental time and the average values for each size were obtained and applied on top of the numerical specimens. Figure \ref{f7}b illustrates the applied boundary condition for the largest specimen at five different average loads $P$.

The crack band width $h_e$ used in the simulations was 2 mm whereas a very fine seed spacing of 0.2 mm had to be used in the direction of crack growth, to capture correctly the stress profile in the ligament. The fracture energy that was indirectly measured from size effect tests was incorporated in the models by calibrating the material point response. Namely, the post-peak response of the two models was calibrated to dissipate approximately the initial fracture energy measured by size effect tests with imposed remote displacements, with $G_f =$ 74.2 N/mm for the SSMM and $G_f =$ 72.1 N/mm for the MTM.

Table \ref{T4} reports, for the coupons of different size, a comparison between the predicted and measured structural strengths. It can be seen that, for both models, predictions agree very well with experiments. Needless to say that had the models been calibrated with the initial fracture obtained by applied remote stresses, the peak loads would be greatly overestimated.

\subsection{Importance of size effect in crashworthiness applications} \label{parametric study}

To clarify the importance of measuring $G_f$ accurately and modeling the fracturing of textile composites correctly, the microplane formulations were used to predict the energy absorbed during the impact of composite crash cans, a situation in which intra-laminar fracture was reported to be the main failure mechanisms. Figure (\ref{f10}c) shows the geometry of the structure under study consisting of a hat section tube and a reinforcing plate glued together by a toughened epoxy glue. The lay-up configuration was $[0^{\circ}]_{11}$ for the hat section tube and $[0^{\circ}]_{8}$ for the plate. The composite tubes, accurately fixed at the bottom, were impacted by a flat mass of 74.4 kg at the velocity of 4.6 m/s in a drop tower.

The crush can was modeled in Abaqus Explicit \cite{Abaqus13} using a mesh of triangular shell elements of 2 mm (see Figure (\ref{f10}d)). All the degrees of freedom of the nodes at the bottom section were fixed while an initial velocity field of 4.6 m/s was prescribed to the impacting mass consisting of rigid shell elements. The general contact algorithm provided by Abaqus Explicit \cite{Abaqus13} was used while element deletion was adopted to avoid excessive element distortion during the simulation. The elements were deleted as soon as the dissipated energy in tension reached $99\%$ of the fracture energy or when the magnitude of the maximum or minimum principal strains reached 0.45.

The comparison between experimental and numerical results is reported in Table \ref{T5} in terms of plateau reaction force on the plate. As can be noted, a very satisfactory agreement is found for both formulations using the fracture energy estimated from size effect. As the table shows, the experimental plateau load, i.e. the reaction force on the plate once the crushing process is stabilized, is 35.6 kN whereas the predicted values are 33.8 kN for SSMM and 32.7 for the MTM. It should be highlighted here that these results represent a pure prediction based on the calibration and validation through uniaxial tests as well as size effect tests only. No adjustment of any of the parameters of the models was done, making the reported results even more remarkable.

Then, to study the importance of $G_f$ on the crashing predictions, simulations were done for the following additional cases: Case 2: fracture energy approximately half of the measured value, and Case 3: fracture energy corresponding to an almost vertical drop in stress after the peak (equal to about 12 N/mm assuming 2 mm as element size). Since, in the latter case, also the shape of the post-peak is predefined, no difference between the two formulations was expected and the simulation was run only with the Spectral Stiffness Microplane Model (Figure \ref{f8}c reports examples of the stress-strain curves for SSMM for all cases). As can be noted from Table \ref{T5} and Figures \ref{f10}a,b, the simulations revealed a huge effect of the fracture energy for both models. In facts, a decrease of the initial fracture energy to about 50$\%$, Case 2, diminished the crushing load to almost a half. Case 3, typical of strength-based constitutive laws implemented in most commercial Finite Element codes, resulted into an underestimation of the crushing load of the order of 70$\%$ for an element size of 2 mm. It is worth remarking that, assuming the stress always drops vertically after the peak, inevitably makes the fracture energy dependent on the element size. Accordingly, the error in Case 3 is mesh dependent and the predicted load decreases with decreasing element size and increases for increasing element size.

\section{Conclusions}

This paper presents and discusses an experimental and numerical investigation on the intra-laminar size effect of textile composites.
Based on the results presented in this study, the following conclusions can be formulated:

\be  \setlength{\itemsep}{0.0mm}

\ii The experimental investigation shows a remarkable size effect in geometrically-scaled textile composite structures failing by intra-laminar fracture propagation. This aspect, too often overlooked in the literature on composites, is the determining factor for damage tolerance design of large composite structures;

\ii The tests agree with Size Effect Law (SEL) proposed by \B~\cite{Baz84,BazPla98}, according to which the fracturing behavior of geometrically scaled structures exhibits a smooth transition from ductile to purely brittle (LEFM type) with increasing structure size;

\ii Size effect tests can be used to determine the fracture characteristics of the composite provided that a) the orthotropic properties of the material are taken into account and b) realistic boundary conditions are assumed for the calculation of the dimensionless energy release rate. The size effect method of measuring the fracture properties is easier to implement than other methods because only peak load measurements are necessary: the post-peak behavior, crack tip displacement measurement and optical measurement of crack tip location are not needed, and even a soft testing machine without servo-control can be used. According to this approach, the initial fracture energy $G_f$ of the investigated carbon twill 2x2 composite was identified to be 73.7 N/mm whereas the effective Fracture Process Zone (FPZ) length, $c_f$ was 1.81 mm. This length is comparable to the size of one tow;

\ii The applicability of SEL to measure the fracture properties of the material was verified numerically by means of two recently proposed microplane models for textile composites \cite{KirSalBaz15a,KirSalBaz15b,SalAshCus15}. Both  formulations matched the size effect data using the $G_f$-value estimated by SEL;

\ii Compared to the experimental results on the axial progressive crushing behavior of composite crush cans, the models calibrated with the measured fracture energy provided an excellent prediction of the crushing load. Further, a parametric study showed that, for both models, a decrease of $G_f$ to about 50$\%$ of the measured value can reduce the crushing load to almost a half. The assumption of a vertical drop of the stress after the peak, typical of strength-based constitutive laws, resulted into an underestimation of the order of 70$\%$ for an element size of 2 mm. The error in this latter case is mesh dependent;

\ii The foregoing results suggest that measuring accurately $G_f$ and modeling correctly the fracturing of textile composites, including their \emph{quasi-brittleness}, is the key in all situations which require accurate prediction of energy absorption (such as crashworthiness analysis) or scaling of mechanical properties. This can be accomplished neither by strength-based approaches, which completely neglect the size effect, nor by LEFM, which does not account for the finiteness of the FPZ.
 \ee

\subsection*{Acknowledgments}

This material is based upon work supported by the Department of Energy under Cooperative Award Number DE-EE0005661 to the United States Automotive Materials Partnership, LLC and sub-award SP0020579 to Northwestern University.
The work was also partially supported under NSF grant No. CMMI-1435923.



\section*{References}
\begin{thebibliography}{00}
\linespread{1}\selectfont
\small




\bibitem[Chou, 1992]{Cho92} Chou T.W. Microstructural Design of Fibre Composites.
Cambridge University Press, New York, 1992.

\bibitem[Daniel and Ishai, 1992]{DanIsh92} Daniel I.M., Ishai O. Engineering Mechanics of
Composite Materials. New York: Oxford University Press, 1992.

\bibitem[Bogdanovich and Pastore, 1996]{BogPas96} Bogdanovich A.E., Pastore C.M. Mechanics of Textile and
Laminated Composites. Chapman and Hall, London, 1996.

\bibitem[\B~ and Planas, (1998)]{BazPla98} \B~Z.P., Planas J., Fracture and Size Effect in Concrete and Other Quasibrittle materials, CRC Press, 1998.

\bibitem[Mollenhauer et al., (2006)]{MolIarKim06} Mollenhauer D, Iarve EV, Kim R, Langley B., Examination of ply cracking in composite laminates with open holes: a Moire´ interferomic and numerical study. Composites – Part A 2006;37: 282–-94.

\bibitem[Green et al., (2007)]{GreWisHal07} Green B.G., Wisnom M.R., Hallet S.R., An experimental investigation into the tensile strength scaling of notched composites, Composites -–
Part A 2007;38:867–-78.

\bibitem[\B~ et al., 1996]{BazDanLi96} \B~Z.P., Daniel I.M., Li Z. Size Effect and Fracture Characteristics of Composite Laminates
 J. Eng. Mater. Technol. 1996;118(3): 317--324.

 \bibitem[Cusatis et al., (2008)]{CusBegBaz08} Cusatis G., Beghini A. \B~Z.P. Spectral Stiffness
Microplane Model for Quasibrittle Composite Laminates—-Part I: Theory J
Appl Mech 2008; 75:0210091--8.

 	

\bibitem[Salviato et al., (2016)]{SalAshCus15} Salviato M., Esna Ashari S., Cusatis G., Spectral stiffness microplane model for damage and fracture of textile composites, Composite Structures 2016;137:170-184.

\bibitem[Kirane et al., 2016]{KirSalBaz15a} Kirane K., Salviato M., \B~Z.P. Microplane triad model for simple and accurate prediction of orthotropic elastic constants of woven fabric composites J Compos Mater, 2016; 50:1247-1260.

\bibitem[Kirane et al., 2016]{KirSalBaz15b} Kirane K., Salviato M., \B~Z.P. Microplane-Triad Model for Elastic and Fracturing Behavior of Woven Composites ASME Journal of Applied Mechanics, 2016; 84:0410061-14.


\bibitem[ASTMD3039, 2014]{ASTMD3039} ASTMD3039 Standard Test Method for Tensile Properties of Polymer Matrix Composite Materials 2014.





\bibitem[\B~ and Tabbara, 1992]{BazTab92} \B~ Z.P., Tabbara M.R. Bifurcation and Stability of Structures
with Interacting Propagating Cracks. Int. J. of Fracture 1992;53:273-289.

\bibitem[ASTM5045, 1999]{ASTM5045} ASTMD5045 Standard Test Methods for Plane-Strain Fracture Toughness and Strain Energy Release Rate of Plastic Materials 1999.




\bibitem[Correlated, 2015]{Correlated} Correlated Solutions, Columbia, USA. http://www.correlatedsolutions.com

\bibitem[Tsai and Wu, 1972]{TsaWu72} Tsai S.W., Wu E.M. A General Theory of Strength for Anisotropic Materials J Compos Mater 1972;5:58--80.


\bibitem[\B~, 1984]{Baz84s} \B~ Z.P., Size Effect in Blunt Fracture: Concrete, Rock, Metal Int. J. Eng. Mech. 1984;110:518-535.


\bibitem[\B and Cedolin, 1991]{BazCed91} Ba\v zant, Z.P., and Cedolin, L. {\em Stability of Structures: Elastic, Inelastic, Fracture and Damage Theories}, Oxford University Press, New York 1991. (2nd ed. Dover Publ.; 3rd ed. World Scientific Publishing, Singapore--New Jersey--London 2010), Section 11.9.

\bibitem[Bao et al., 1992]{Bao91} Bao G., Ho S., Suo Z., Fan B. The Role of Material Orthotropy in Fracture Specimens for Composites
 Int J Solid Structures 1992;29: 1105-1116 and Corrigenda: Int J Solid Structures 1992;29: 2115.

 \bibitem[Abaqus, 2013]{Abaqus13} ABAQUS, v., 2013. ABAQUS User‘s Manual, Ver. 6.13-1. Hibbit, Karlson and Sorenson, Pawtucket, RI.

\bibitem[Barsoum, 1974]{Barsoum74} Barsoum R.S. Application of Quadratic Isoparametric Finite Elements in Linear Fracture Mechanics.
Int. J. Fracture 1974;10:603–-605.

\bibitem[Rice, 1968]{Rice68} Rice, J.R. A path independent integral and the approximate analysis of strain concentrations by notches
and cracks. J. Appl Mech ASME 1968;35:379–-386.



 \bibitem[\B~and Oh, 1983]{BazOh83} \B~Z.P., Oh B.H. Crack band theory for fracture of concrete. Matériaux et construction 1983;16(3):155--177.


\bibitem{Baz84} Ba\v zant, Z.P. Microplane model for strain-controlled inelastic behavior. Chapter 3 in {\em ``Mechanics of Engineering Materials}." ed. by C. S. Desai and R. H. Gallagher, J. Wiley, London, 1984; 45--59.

\bibitem[\B~ and Oh, 1985]{BazOh85} \B~Z.P., Oh B. H. Microplane model for progressive fracture of concrete and rock, J. Eng. Mech., ASCE, 1985; 111:559--582.



\bibitem[Caner and \B, 2013]{CanBaz13} Caner F.C., \B~Z.P. Microplane Model M7 for Plain
Concrete. I: Formulation J. Eng. Mech., ASCE, 2013;139:1714--1723.


\bibitem[Cusatis and Schauffert, (2009)]{CusSch09} Cusatis G., Schauffert E. A. Cohesive crack analysis of size effect. Eng Fract Mech 2009;76:2163--2173.

\bibitem{HooBaz14} Hoover, C.G., and Ba\v zant, Z.P. (2014). ``Cohesive crack, size effect, crack band and work-of-fracture models compared to comprehensive concrete fracture tests." {\em Int. J. of Fracture} 187 (1), pp. 133-143.

\end{thebibliography}


\section*{Appendix}\label{appendix}
Table \ref{T6-1} and Table \ref{T6-2} present the calibrated parameters for Spectral Stiffness Microplane Model (SSMM) and Microplane Triad Model (MTM), respectively according to Refs. \cite{SalAshCus15, KirSalBaz15a,KirSalBaz15b}.

\hfil
\hfil
\hfil
\begin{table}[ht]
\centering  
\begin{tabular}{l c c} 
 \hline
  \rule{0pt}{4ex}
 Description & Symbol (units) & Measured value\\[1 ex]
 \hline
Fiber volume fraction & $V_f$ (-) & 0.54\\
Laminate thickness & $t$ (mm) & 1.9\\
In-plane modulus & $E$=$E_{1}$=$E_{2}$ (GPa)  & 53.5\\
In-plane shear modulus & $G$ = $G_{12}$ (GPa)  & 4.5 \\
In-plane Poisson ratio & $\nu=$$\nu_{12}$ = $\nu_{32}$ (-) & 0.055  \\
In-plane tensile strength in direction 1 and 2 & $F_{1t}$ = $F_{2t}$ (MPa) & 598 \\
\hline
\end{tabular}
\caption{\sf Properties of carbon twill 2x2/epoxy composite}
\label{T1}
\end{table}
\clearpage
\begin{table}[ht]
\centering  
\begin{tabular}{l c c c c c}
\hline
\rule{0pt}{4ex}Size & Width, &Gauge length,  &Length, &Crack length,  &Thickness, \\
&$D$&$L$&$L=L+2L_t$&$a_0$&$t$\\[1 ex]
\hline
Small & $20$ & 44.5 & 120.5 & 4 & 1.9\\
Medium & $40$& 89.0 & 165.0 & 8 & 1.9\\
Large & $80$ & 178.0& 254.0 & 16& 1.9 \\
\hline
\multicolumn{6}{l}{Units: mm. Tab length $L_t=$ 38 mm for all investigated sizes.}
\end{tabular}
\caption{\sf Geometrical specifications of the SENT specimens under study}
\label{T2}
\end{table}

\begin{table}[ht]
\centering  
\begin{tabular}{c c c c}
\hline
\rule{0pt}{4ex}Specimen gauge length,  & Specimen width, &Max, load &Nominal strength\\
$L$ (mm)& $D$ (mm)& $P_{\mbox{max}}$ (kN)& $\sigma_N$ (MPa)\\[1 ex]
\hline
\multirow{3}{*}{44.5}& \multirow{3}{*}{20}& 13.17 & 350.27 \\
 &                                          & 12.67 & 336.81 \\
 &                                          & 13.42 & 356.91 \\
\hline
\multirow{3}{*}{89.0}& \multirow{3}{*}{40}& 16.58 & 220.48 \\
&                                          & 18.69 & 248.54 \\
&                                       & 19.83 & 263.70 \\
\hline
\multirow{3}{*}{178.0}& \multirow{3}{*}{80} & 30.17 & 200.60 \\
&                                             & 29.60 & 190.81 \\
&                                            & 29.65 & 197.14 \\
\hline
\end{tabular}
\caption{\sf Results of tensile tests on Single End Notched Specimens.}
\label{T3}
\end{table}

\begin{table}[ht]
\centering  
\begin{tabular}{l c c c c }
\hline
\rule{0pt}{4ex}\multirow{2}{*}{Size}&\multirow{2}{*}{Width (mm)}& \multicolumn{3}{c}{Structural strength, $\sigma_N$ (MPa)}\\
\cline{3-5}
& &Experiments& SSMM \cite{SalAshCus15}& MTM \cite{KirSalBaz15b}  \\
\hline
Small & $20$ & 348.0 $\pm 10.2$ & 345.2 & 345.6 \\
Medium & $40$& 224.2 $\pm 21.9$ & 255.5 & 258.8 \\
Large & $80$ & 196.2 $\pm 5.0$& 199.3 & 198.1\\
\hline
\end{tabular}
\caption{\sf Comparison between experimental and predicted nominal strength of the SENT specimens under investigation.}
\label{T4}
\end{table}

\begin{table}[ht]
\centering  
\begin{tabular}{l c c c c c c c}
\hline
\rule{0pt}{4ex}\multirow{2}{*}{Description (units)}&\multirow{2}{*}{Exp.~}&\multirow{2}{*}{case $\#$} &\multicolumn{2}{c}{SSMM \cite{SalAshCus15}}&& \multicolumn{2}{c}{MTM \cite{KirSalBaz15b}}\\
\cline{4-5}\cline{7-8}
& & & value& $\Delta\%$ & & value& $\Delta\%$ \\
\hline
\multirow{3}{*}{Plateau load (kN)}& \multirow{3}{*}{$35.6$} & 1& 33.8& 5.06 && 32.7& 8.15\\
 &  & 2& 20.2 & 43.26& & 20.9& 41.29\\
 &  & 3& 9.0& 74.72 && /& /\\
\hline
\end{tabular}
\caption{\sf Experimental and predicted plateau crushing load for various value of the initial fracture energy. Case 1: $G_{f}^{(1)}=73.7$ N/mm (as estimated by SEL, Eq. (\ref{e11})); Case 2: $G_{f}^{(2)}\approx 1/2 ~G_{f}^{(1)}$; Case 3: $G_{f}^{(3)}\approx$ 12 N/mm (corresponding to almost vertical drop in stress after peak).  }
\label{T5}
\end{table}

\begin{table}[ht]
\centering  
\resizebox{\columnwidth}{!}{%
\begin{tabular}{c l c c} 
 \hline
  \rule{0pt}{4ex}
 Mode &Description & Symbol (units) & Calibrated value\\[1 ex]
 \hline
\multirow{8}{*} {12} & mode 1 elastic eigenvalue & $\lambda^{(1)}$ (GPa) & 61.85$$\\
& mode 2 elastic eigenvalue & $\lambda^{(2)}$ (GPa) & 50.71$$\\
& microplane peak stress in tension & $s_0^{(12)}$ (MPa) & 400\\
& parameter governing post-peak softening in tension  & $k_{bt}^{(12)}$ (-) & $30.60\times10^{-3}$\\
& parameter governing post-peak softening in tension  & $a_{12t}$ (-) & 0.75\\
& microplane peak stress in compression & $c_0^{(12)}$ (MPa) & 405\\
& parameter governing post-peak softening in compression  & $k_{bc}^{(12)}$ (-) & $30.60\times10^{-3}$\\
& parameter governing post-peak softening in compression  & $a_{c12}$ (-) & 0.75\\[1 ex]
\hline
\multirow{8}{*} {4} & mode 4 elastic eigenvalue & $\lambda^{(4)}$ (GPa) & 8.10$$\\
& microplane stress in tension at start of non-linear boundary & $s_0^{(4)}$ (MPa) & 45\\
& exponent governing pre-peak non-linearity in tension and compression  & $p$ (-) & $0.3$\\
& strain at starting of post-peak softening in tension  & $k_{at}^{(4)}$ (-) & $124.6\times10^{-3}$\\
& parameter governing post-peak softening in tension  & $k_{bt}^{(4)}$ (-) & $120.15\times10^{-3}$\\
& microplane stress in compression at start of non-linear boundary & $c_0^{(4)}$ (MPa) & 45\\
& strain at starting of post-peak softening in compression  & $k_{ac}^{(4)}$ (-) & $124.6\times10^{-3}$\\
& parameter governing post-peak softening in compression  & $k_{bc}^{(4)}$ (-) & $120.15\times10^{-3}$\\[1 ex]
\hline
\multirow{7}{*} {3} & mode 3 elastic eigenvalue & $\lambda^{(3)}$ (GPa) & 10.82$$\\
& microplane peak stress in tension & $s_0^{(3)}$ (MPa) & 90\\
& strain at starting of post-peak softening in tension  & $k_{at}^{(3)}$ (-) & $4.0\times10^{-3}$\\
& parameter governing post-peak softening in tension  & $k_{bt}^{(3)}$ (-) & $20\times10^{-3}$\\
& microplane peak stress in compression & $c_0^{(3)}$ (MPa) & 90\\
& strain at starting of post-peak softening in compression  & $k_{ac}^{(3)}$ (-) & $4.0\times10^{-3}$\\
& parameter governing post-peak softening in compression  & $k_{bc}^{(3)}$ (-) & $20\times10^{-3}$\\[1 ex]
\hline
\multirow{8}{*} {5} & mode 5 elastic eigenvalue & $\lambda^{(5)}$ (GPa) & 7.20$$\\
& microplane stress in tension at start of non-linear boundary & $s_0^{(5)}$ (MPa) & 45\\
& exponent governing pre-peak non-linearity in tension and compression  & $p$ (-) & $0.3$\\
& strain at starting of post-peak softening in tension  & $k_{at}^{(5)}$ (-) & $124.6\times10^{-3}$\\
& parameter governing post-peak softening in tension  & $k_{bt}^{(5)}$ (-) & $120.15\times10^{-3}$\\
& microplane stress in compression at start of non-linear boundary & $c_0^{(5)}$ (MPa) & 45\\
& strain at starting of post-peak softening in compression  & $k_{ac}^{(5)}$ (-) & $124.6\times10^{-3}$\\
& parameter governing post-peak softening in compression  & $k_{bc}^{(5)}$ (-) & $120.15\times10^{-3}$\\[1 ex]
\hline

\end{tabular}
}
\caption{\sf Calibrated parameters of Spectral Stiffness Microplane Model (SSMM).}
\label{T6-1}
\end{table}

\begin{table}[ht]
\centering  
\resizebox{\columnwidth}{!}{%
\begin{tabular}{l c c} 
 \hline
  \rule{0pt}{4ex} Description & Symbol (units) & Calibrated value\\[1 ex]
 \hline
 axial and transverse modulus for matrix&$E^m$(GPa)&3.5\\
 axial modulus for fiber&$E^f_{1'}$(GPa)&190\\
 axial modulus for yarn&$E^y_{1'}$(GPa)&134\\
 transverse modulus for fiber&$E^f_{2'}$ = $E^f_{3'}$(GPa)&40\\
 transverse modulus for yarn&$E^y_{2'}$ = $E^y_{3'}$(GPa)&9.69\\
 shear modulus for matrix&$G^m$(GPa)&1.7\\
 in-plane shear modulus for fiber&$G^f_{1'2'}$ = $G^f_{1'3'}$(GPa)&24\\
 in-plane shear modulus for yarn&$G^y_{1'2'}$ = $G^y_{1'3'}$(GPa)&5.68\\
 out-plane shear modulus for fiber&$G^f_{2'3'}$(GPa)&14.3\\
 out-plane shear modulus for yarn&$G^y_{2'3'}$(GPa)&3.57\\
 Poisson's ratio for matrix&$\nu^m$&0.35\\
 in-plane Poisson's ratio for fiber&$\nu^f_{1'2'}$ = $\nu^f_{1'3'}$&0.26\\
 in-plane Poisson's ratio for yarn&$\nu^y_{1'2'}$ = $\nu^y_{1'3'}$&0.287\\
 out-plane Poisson's ratio for fiber&$\nu^f_{2'3'}$&0.49\\
 out-plane Poisson's ratio for yarn&$\nu^y_{2'3'}$&0.448\\
 microplane normal strain at which initial drop in load occurs (T) & $\varepsilon_N^{f0}$ & 0.0109$$\\
 microplane normal strain at which initial drop in load occurs (C) & $\varepsilon_N^{f0c}$ & 0.011227$$\\
 microplane normal strain at which progressive softening begins & $\eps_N^{f1}$ & 0.014933\\
 parameter governing the sharpness of the initial drop in stress & $R$ & 0.6216\\
 parameter governing the shape of the progressive fiber damage & $R_1$ & 0.0082\\
 parameter governing the shape of the progressive fiber damage & $R_2$ & 160,000\\
 parameter governing the shape of the progressive fiber damage & $q$ & 4.25\\
 effective strain at which pre-peak non-linearity begins in the matrix & $\varepsilon^{m0}$ & 0.007\\
 effective strain at peak load in the matrix & $\varepsilon^{mc}$ & 0.22\\
 parameter governing the shape of the non-linear arm & $H$ & 92.1\\
 parameter governing the shape of the non-linear arm & $n$ & 0.5714 \\
 effective strain at complete failure in the matrix & $\varepsilon^{mf}$ & 0.35\\[1 ex]
\hline

\end{tabular}
}
\caption{\sf Calibrated parameters of Microplane Triad Model (MTM).}
\label{T6-2}
\end{table}

\clearpage

\bfi \center
  \includegraphics[width = 0.6\textwidth]{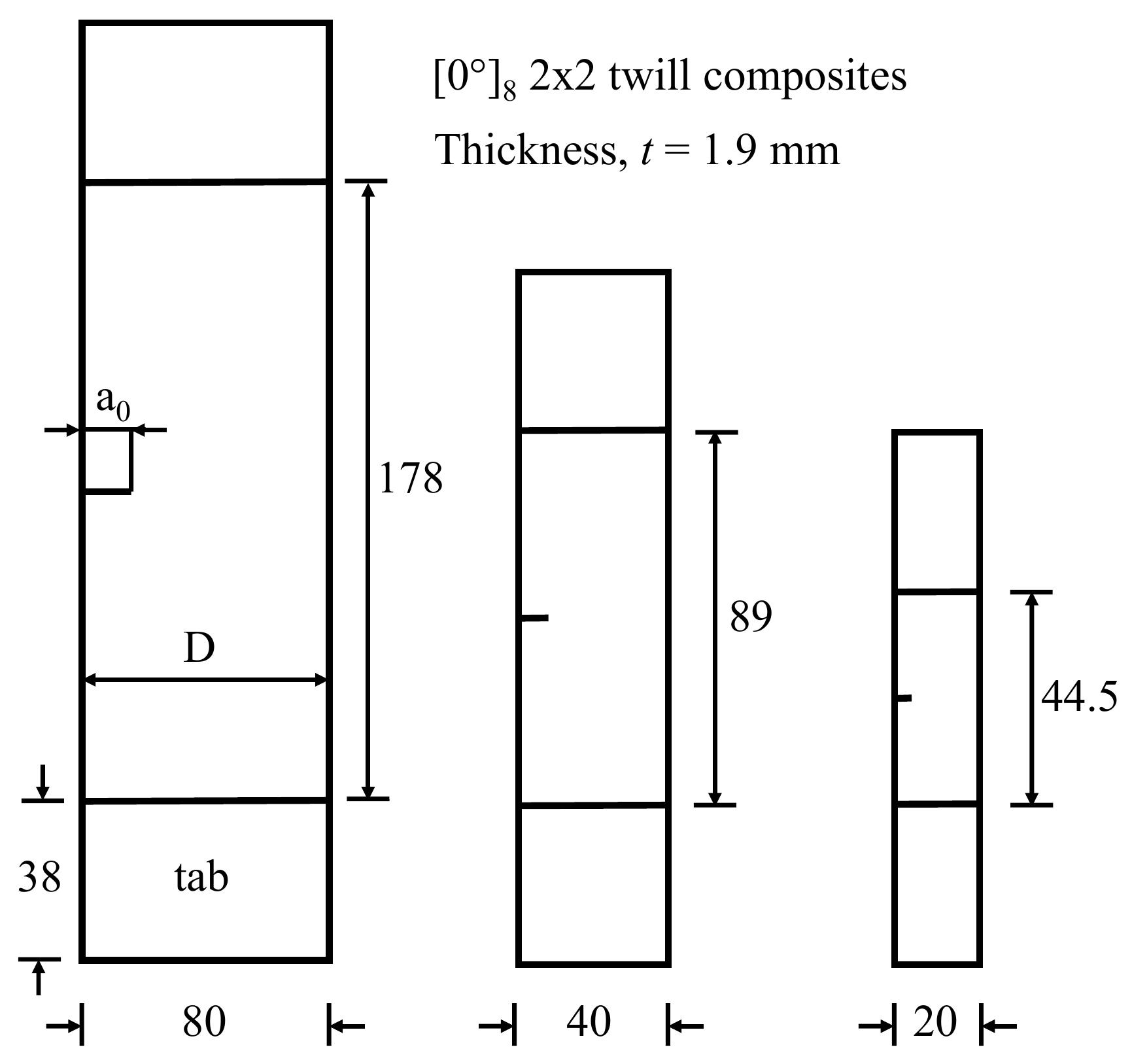} \caption{\label{f1} \sf Geometry of Single Edge Notch Tension (SENT) specimens under study. Units: mm.} \efi
\bfi \center
  \includegraphics[width = 0.6\textwidth]{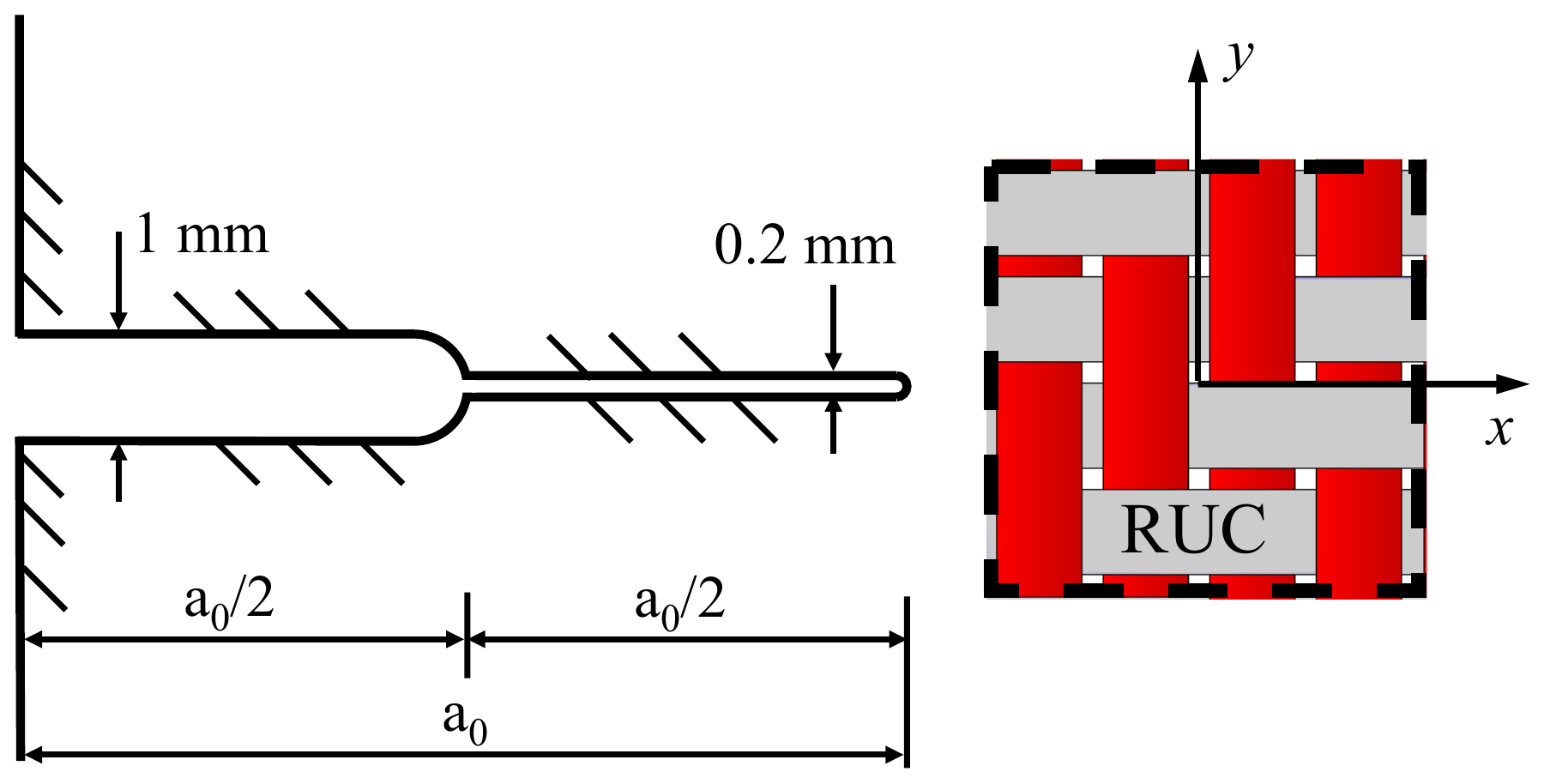} \caption{\label{f2} \sf Geometry of the notch (scaled for each size) and schematic of the Representative Unit Cell of the twill 2x2 composites under study.} \efi

\bfi \center
  \includegraphics[width = 1.0\textwidth]{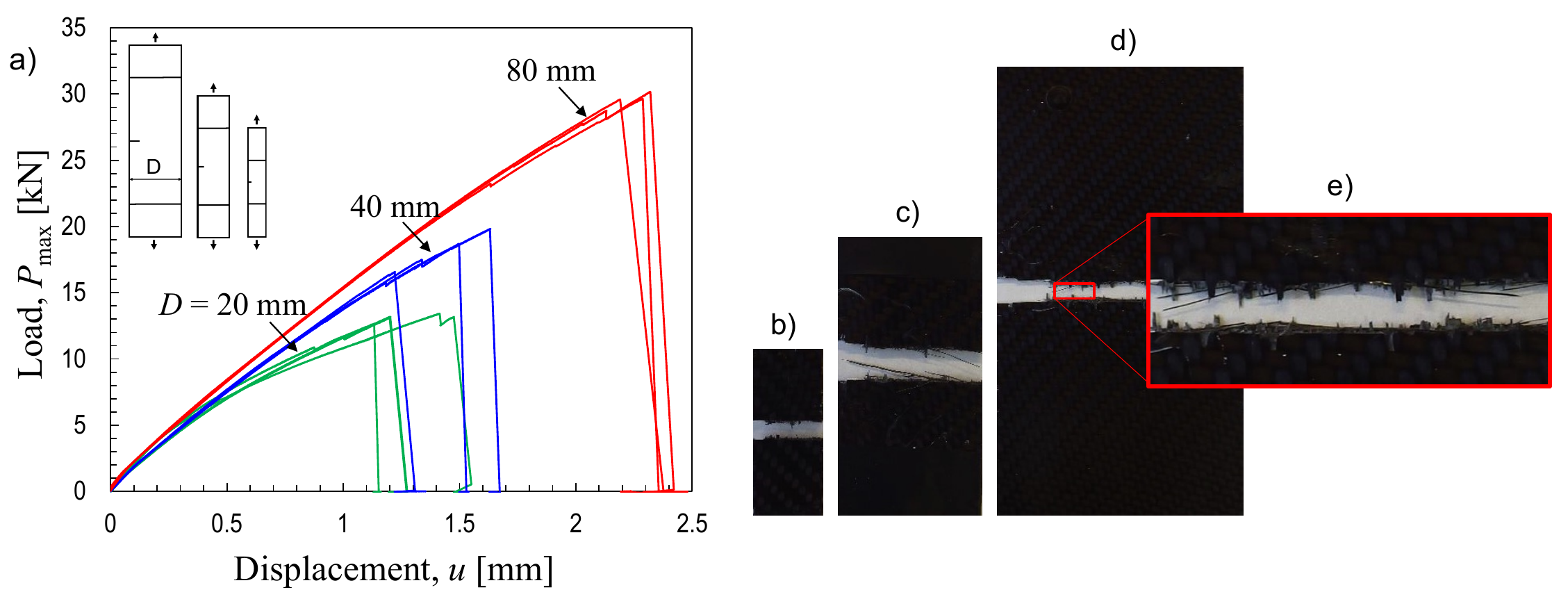} \caption{\label{f3} \sf a) Typical load-displacement curves of $[0^{\circ}]_8$ geometrically-scaled SENT specimens of various sizes, showing decreasing nonlinearity increasing specimen dimensions. Typical failure patterns of Single Edge Notched specimens for width b) $D=20$ mm, c) $D=40$ mm and d) $D=80$ mm. e) Magnification of fracture surface for the large size specimen showing extensive tow failure and pull-out.} \efi
\bfi \center
  \includegraphics[width = 1.0\textwidth]{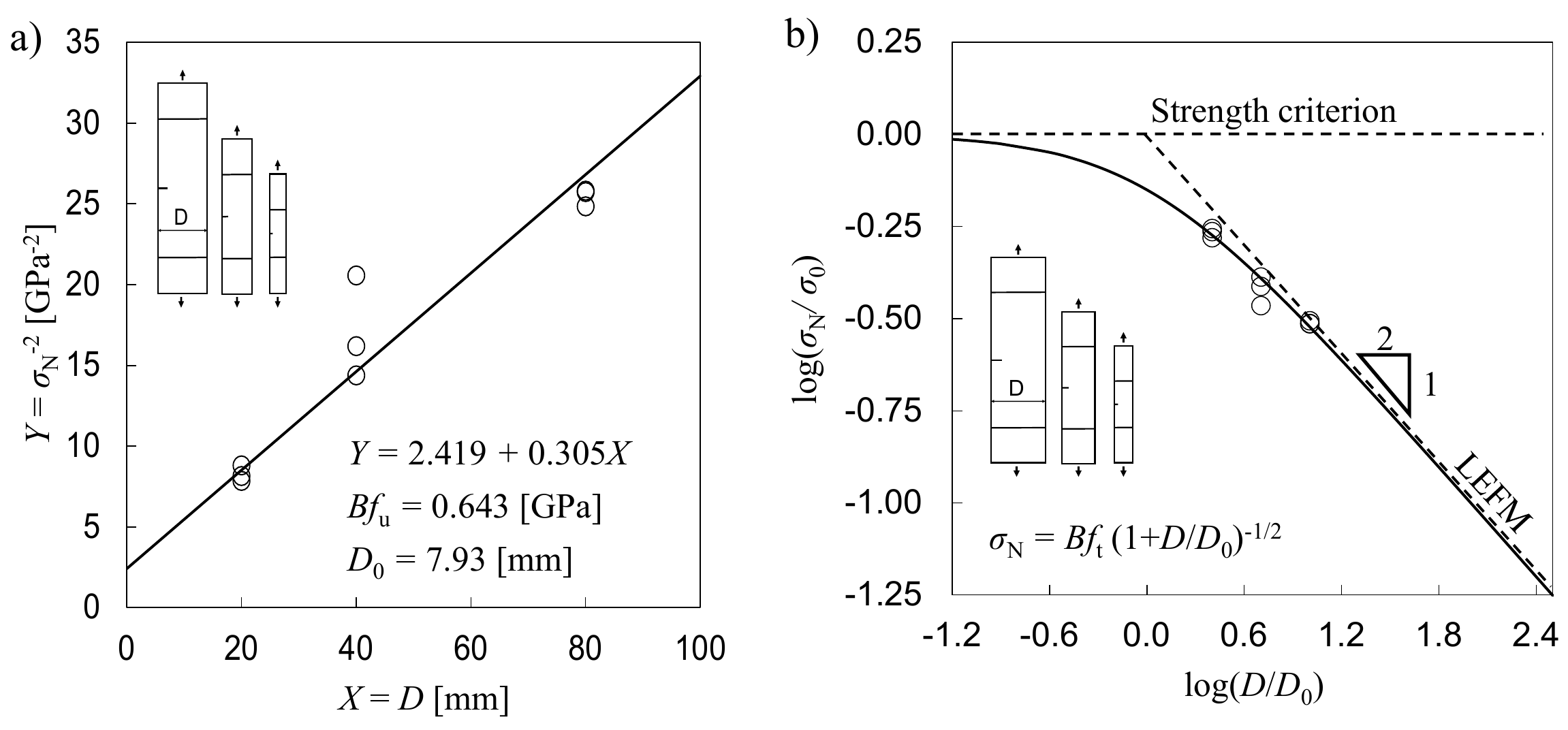} \caption{\label{f5} \sf Size effect study. a) Linear regression analysis to characterize the size effect parameters. b) Measured size effect for $[0^{\circ}]_8$ twill 2x2 laminates. } \efi

\bfi \center
  \includegraphics[width = 0.8\textwidth]{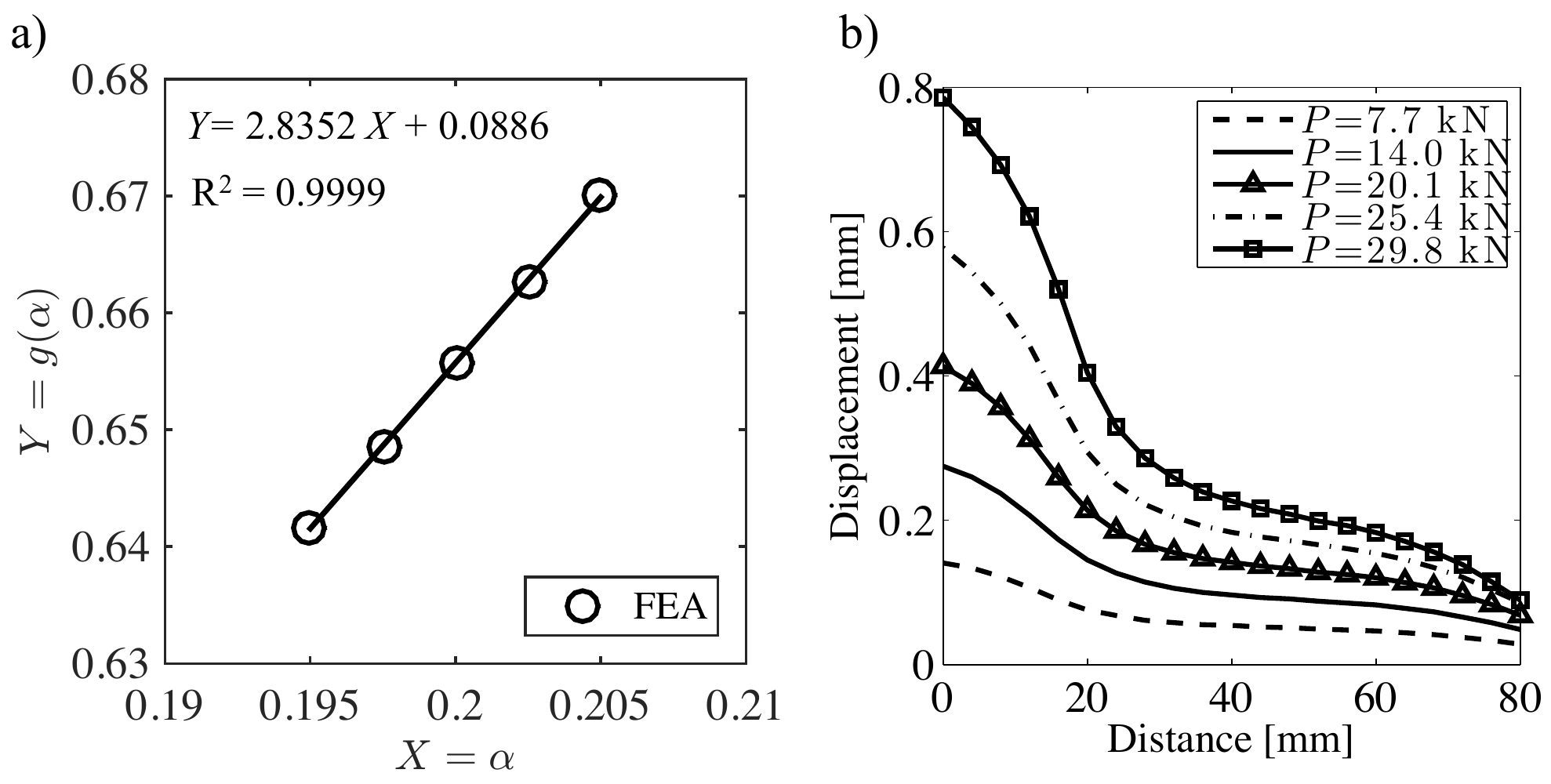} \caption{\label{f7} \sf a) Calculation of the dimensionless energy release rate $g(\alpha_0)$ by linear interpolation of FEA. b) Applied boundary conditions in FEA simulations of the largest size specimen for different average loads $P$.} \efi
%
\bfi \center
  \includegraphics[width = 0.88\textwidth]{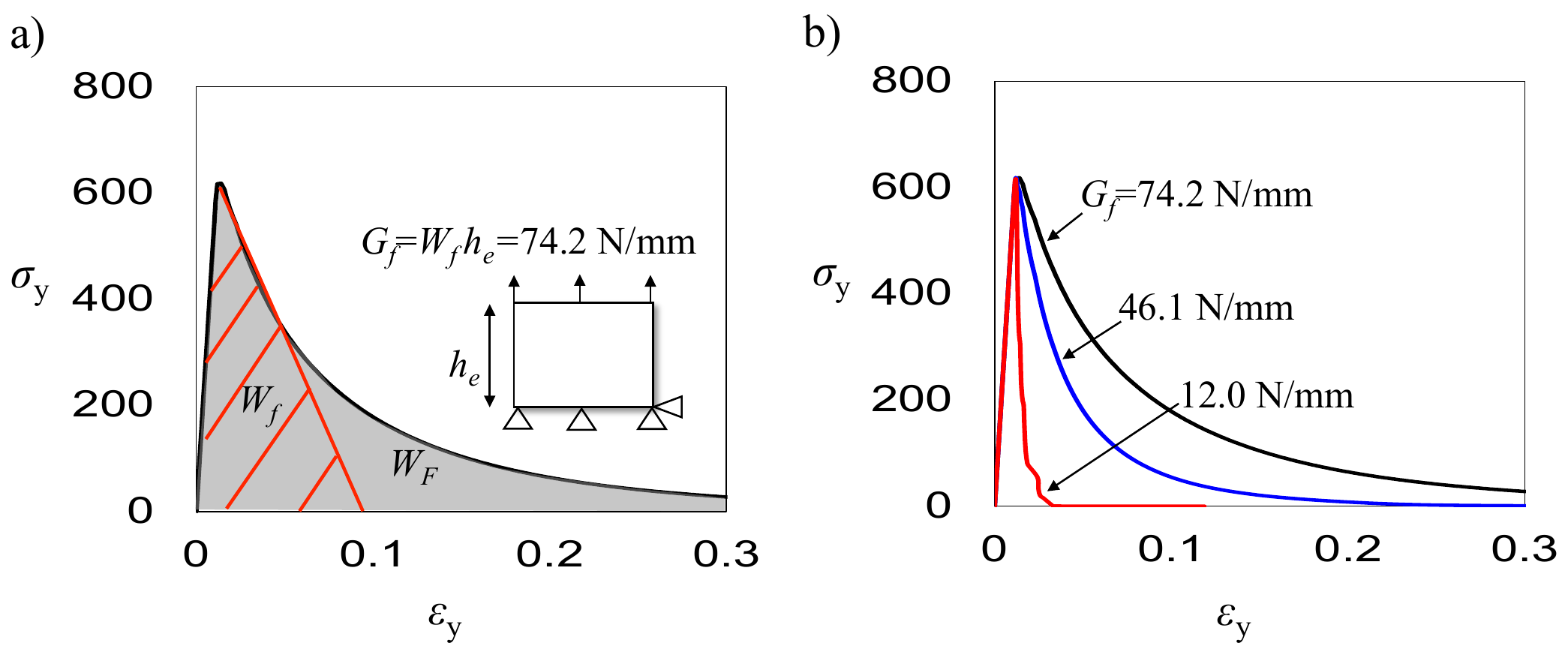} \caption{\label{f8} \sf a) Typical stress-strain curve in pure tension for the Spectral Stiffness Microplane Model \cite{SalAshCus15}. The initial fracture energy $G_f$ is calibrated adjusting the post-peak softening response of the material. b) Typical stress-strain curves in tension provided by \cite{SalAshCus15} for different values of the initial fracture energy: 1) $G_f$ = 73.7 N/mm, 2) $G_f$ = 46.1 N/mm and 3) $G_f$ = 12.0 N/mm (corresponding to almost vertical drop of stress after the peak).} \efi
\bfi \center
  \includegraphics[width = 1.0\textwidth]{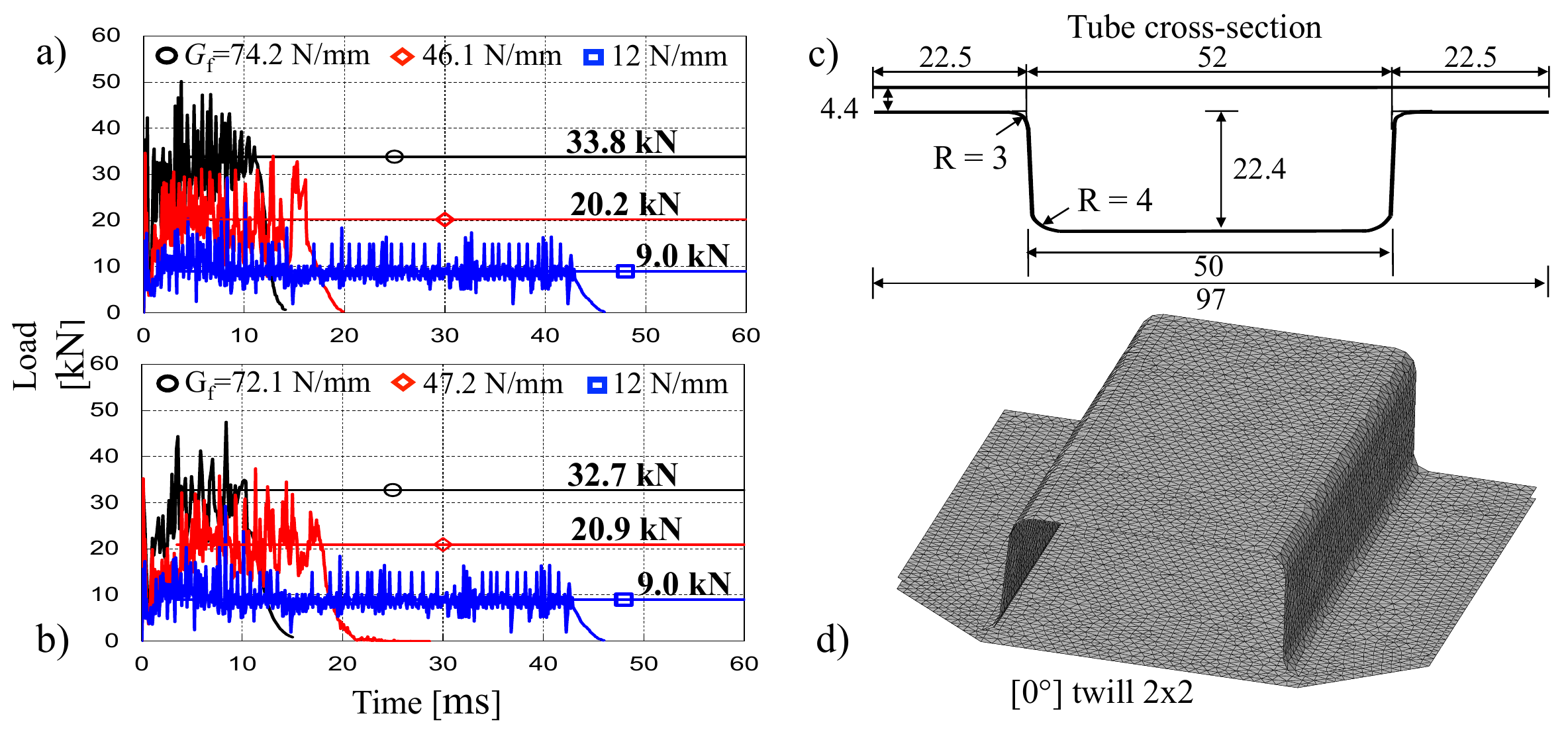} \caption{\label{f10} \sf Crashing of composite tubes. (hat section: $[0^{\circ}]_{11}$, plate: $[0^{\circ}]_8$). a) Crashing load vs time predicted by Spectral Stiffness Microplane Model \cite{SalAshCus15} for various values of intra-laminar fracture energy, $G_f$; b) Crashing load vs time predicted by Microplane Triad Model \cite{KirSalBaz15a,KirSalBaz15b} for various values of intra-laminar fracture energy, $G_f$; c) geometric specifications of the crash can cross-section (dimensions in mm); d) Typical FE mesh used in the simulations. } \efi

%
%

\end{document}